\documentclass[aps,pra,superscriptaddress,12pt,tightenlines,nofootinbib]{revtex4}
\usepackage{amsmath,amsthm,graphicx,amssymb,verbatim,listings}
\usepackage{wasysym}
\usepackage[T1]{fontenc}     
\usepackage{lmodern}         
\usepackage[ pdftex, plainpages = false, pdfpagelabels,
                 pdfpagelayout = useoutlines,
                 bookmarks,
                 bookmarksopen = true,
                 bookmarksnumbered = true,
                 breaklinks = true,
                 linktocpage=all,
                 pagebackref=false,
                 colorlinks = true,
                 linkcolor = BrickRed,
                 urlcolor  = blue,
                 citecolor = BrickRed,
                 anchorcolor = green,
                 hyperindex = true,
                 hyperfigures
                 ]{hyperref}
\usepackage[usenames, dvipsnames]{xcolor}
\usepackage{xifthen} 

\newcommand{\arxiv}[2][]{\ifthenelse{\isempty{#1}}{\href{http://arxiv.org/abs/#2}{{\tt arXiv:\allowbreak{}#2}}} {\href{http://arxiv.org/abs/#2}{{\tt arXiv:\allowbreak{}#2 [#1]}}}}

\newcommand{\booktitle}{\textsl}
\newcommand{\hrefdoi}[2]{\href{https://dx.doi.org/#1}{#2}}

\begin{document}

\title{An Underappreciated Exchange in the Bohr--Einstein Debate}

\author{Blake C.\ Stacey}
\affiliation{\href{http://www.physics.umb.edu/Research/QBism/}{Physics
    Department}, University of Massachusetts Boston}

\date{\today}

\maketitle

The Bohr--Einstein debate is one of the more remarkable protracted
intellectual exchanges in the history of physics.  Its influence has
been lasting:\ One of the few clear patterns in a 2013 survey about
quantum foundations~\cite{Schlosshauer:2013} was that the physicists
who believed Bohr to be correct were apt to say that Einstein had been
wrong. The exchanges began when Bohr and Einstein first met in 1920,
continued at the Solvay conferences of the following decade, reached a
dramatic crescendo with the EPR paradox in 1935, and continued
thereafter~\cite{Bohr:1949}. Not every episode in this long story has
been investigated equally~\cite{Chevalley:1999}. In particular, one
late statement attributed to Bohr has received much more intense
examination than Einstein's equally pithy reply.

It is often said that Bohr declared~\cite{Petersen:1963},
\begin{quote}
  There is no quantum world. There is only an abstract quantum
  physical description. It is wrong to think that the task of physics
  is to find out how nature is. Physics concerns what we can say about
  nature.
\end{quote}
Philosophers of science have read this and marveled at the decadent,
positivistic indulgence, at the almost solipsistic temperament!  They
see this and maintain that for Bohr, the quantum is the closing of the
book of nature. The great pursuit of objective truth is --- so it would
seem --- all but abandoned, replaced by mere instrumentalism, by the
reduction of physics to a mental device for adequate predictions.

The philosophy-of-science literature has considered this particular
Bohr-ism in some depth (see \cite{Mermin:2004, Fuchs:2018}, for
example), while it has mostly neglected Einstein's direct response to
it.  The reason for this disparity must remain a matter of conjecture,
but one plausible contributing factor is that the exchange was not put
in print until well after the heyday of the debate. In any event, we
now turn to Einstein's reply so that we may begin filling this
lacuna. Einstein riposted,
\begin{quote}
  There is no quantum world. There is only an abstract quantum
  physical description.
\end{quote}
One can almost hear the soft and softly-accented voice of the flawed
but kindly man, displaced in an unfathomably harsh world. \emph{There
  is no quantum world} --- of course, Einstein has been insistent that
quantum mechanics is incomplete. The stochasticity of detector clicks
is not the spontaneity of the Old One, but only our ignorance of the
true physical configuration. \emph{There is only an abstract quantum
  physical description} --- for it is not the world that is quantum,
but only our approximate and unfinished description of it. And, having
thus reminded us of what he feels so keenly, the separation between
the ``laws'' of \emph{physics} and the laws of \emph{nature,} Einstein
strikes a note that is a little chiding, a little mournful:
\begin{quote}
  It is wrong to think that the task of physics is to find out how
  nature is. Physics concerns what we can say about nature.
\end{quote}
Physics is a human activity, a collective enterprise by a species
equipped with shortcomings he knew only too well. Einstein, who could
be quite critical of how physicists picked their topics of
concern~\cite{OC:2017}, reiterates that dissatisfaction
here. \emph{Physics concerns} --- Einstein is not the man to use that
turn of phrase with wholehearted approval. We may be content today
with a successful but incomplete theory; we should not remain so
forever.

Quotations are often shared without careful regard for their origins,
but the attribution of these lines to Bohr and to Einstein are equally
solid~\cite{Mermin:2004}.  It is mildly inconvenient that the passages
from Bohr and Einstein are verbally identical, but this difficulty
pales into insignificance against the interplay of the contrasting
visions they express.

\end{document}